\documentclass[11pt]{article}
\pdfoutput=1
\usepackage{jheppub}
\usepackage[usenames,dvipsnames,table]{xcolor}
\usepackage{graphicx,amsmath,amssymb,amsthm,multirow,array}

\usepackage[numbers,sort&compress]{natbib}
\usepackage{subcaption}
\usepackage{float}
\usepackage{afterpage}
\definecolor{rust}{rgb}{0.8,0.2,0.2}

\def\AdS#1{AdS$_{#1}$}
\def\amp{\alpha}

\def\IZ{\relax\ifmmode\hbox{Z\kern-.4em Z}\else{Z\kern-.4em Z}\fi}

\def\vT{{\mathfrak v}}
\def\GT{G_{\scriptsize T}}
\def\amp{{\mathfrak a}}

\def\bi{\begin{itemize}} \def\ei{\end{itemize}}

\title{Cosmological particle production at strong coupling}

\author[a]{Mukund Rangamani}
\author[b]{\!\!, Moshe Rozali}
\author[b]{\!\!, Mark Van Raamsdonk}

\affiliation[\,a]{Centre for Particle Theory \& Department of Mathematical Sciences,\\
Durham University, South Road, Durham DH1 3LE, UK.}
\affiliation[\,b]{Department of Physics and Astronomy, University of British Columbia,\\
6224 Agricultural Road, Vancouver, BC V6T 1Z1, Canada.}
\emailAdd{mukund.rangamani@durham.ac.uk}
\emailAdd{rozali@phas.ubc.ca}
\emailAdd{mav@phas.ubc.ca}

\abstract
{ We study the dynamics of a strongly-coupled quantum field theory in a cosmological spacetime using the holographic AdS/CFT correspondence. Specifically we consider a confining gauge theory in an expanding FRW universe and track the evolution of the stress-energy tensor during a period of expansion, varying the initial temperature as well as the rate and amplitude of the expansion. At strong coupling, particle production is inseparable from entropy production. Consequently, we find significant qualitative differences from the weak coupling results: at strong coupling the system rapidly loses memory of its initial state as the amplitude is increased. Furthermore,  in the regime where the
Hubble parameter is much smaller than the initial temperature, the dynamics is well-modelled as a plasma evolving hydrodynamically.}

\begin{document}
\maketitle

\section{Introduction}
\label{sec:intro}

The dynamics of quantum fields in curved spacetime is of immense physical importance in our universe.
Quantum fluctuations and particle production provide the seeds for the macroscopic structures we
observe in the present epoch. While the standard inflationary paradigm works extremely well to explain the observations to date, it is based on the dynamics of free fields. On the other hand, it is conceivable that strongly-coupled field theory dynamics played a significant role during some phase of the very early universe. This motivates the study of strongly-coupled QFTs in cosmological spacetimes.

Such an investigation would be impractical using direct field theory methods, which at best can quantitatively answer only very simple questions when strong interactions are present. However, if the field theory in question is holographic -- i.e., is dual via the AdS/CFT correspondence to a gravitational theory in an asymptotically Anti-de-Sitter spacetime -- we can translate questions about strongly-coupled cosmological physics into much simpler questions in classical gravity. This approach provides a powerful quantitative insight into dynamics of strongly-coupled QFTs in curved spacetime; cf., \cite{Marolf:2013ioa} for a review.

In this paper, we focus on a classic question in the study of quantum fields on curved spacetime. Starting from a thermal equilibrium state in Minkowski space, we investigate the final state of the field theory after a period of homogeneous and isotropic expansion (see \cite{Chesler:2008hg} for other scenarios). In general the evolution entails time-dependent couplings in the field theory which lead to energy and entropy  production. We are interested in ascertaining if there are qualitative differences in this production when compared with results in weakly-coupled field theory, where it can be understood as particle production in a time-dependent background. Roughly speaking, at strong coupling particle production is expected to be accompanied by rapid local thermalization, a feature that is postponed at weak coupling to the end of the expansion. We will see that this lack of separation of time scales in the strongly coupled theory leads to qualitatively different results from the weakly coupled context.

The simplest field theories to study holographically are certain strongly-coupled conformal field theories (CFTs) with many degrees of freedom. 
However, a CFT on an FRW cosmological spacetime does not have particle/energy production; a FRW spacetime is related by a conformal transformation to a static universe.\footnote{ Of course in making this assertion we are eliding over the fact that for even spacetime dimensions, CFTs have conformal anomalies. The conformal anomaly  leads to  a coupling independent Casimir type energy which is simply given by the local scale factor. It does not thus capture the physical effect of particle production we are after. Note however that it does have physical consequences for discussion of inflationary physics, cf.,  \cite{Hawking:2000bb}.} We therefore consider a non-conformal QFT$_{3+1}$ obtained by compactifying a CFT$_{4+1}$ on a circle. Studying this QFT$_{3+1}$ on an FRW spacetime is equivalent to studying the CFT$_{4+1}$ on a spacetime ${\cal B} = \text{FRW} \times {\bf S}^1$.

Using the holographic duality, this translates to a classical gravity question of finding asymptotically \AdS{6} spacetimes whose boundary geometry is ${\cal B}$. By solving the relevant partial differential equations numerically, we can find solutions corresponding to a given initial temperature and FRW scale factor $a(t)$. The physical data of the final state (thermodynamic and otherwise) after the expansion can be read off from the solution.

Using these methods, we study how the amount of energy/entropy production are related to the rate and duration of cosmological expansion. Comparing with free field theory, we find interesting differences. As the amplitude is increased, the amount of energy produced during the expansion quickly becomes independent of the initial temperature, a result not seen in the free field limit. Further, at strong coupling when the maximum Hubble parameter is much smaller than the scale set by the temperature, the results for the temperature change show excellent agreement with analytic results from a hydrodynamic model, providing a strong check of the numerical results.

\section{Generalities}
\label{sec:general}

We consider a holographic CFT$_{4+1}$ on the spacetime
\begin{equation}
\label{FRW}
ds^2 = -d t^2 + dx_c^2 + a^2(t)\,  d\vec{x}^2 \,,
\end{equation}
where $x_c$ parameterizes a ${\bf S}^1$ of size $\ell_c$, and anti-periodic boundary conditions are imposed for fermions. The scale factor $a(t)$  evolves from $a=1$ in the limit $t = -\infty$ to $a=\amp$ at $t = \infty$. For explicit calculations, we  choose
\begin{equation}
\label{timedep}
a(t) = {\amp + 1 \over 2} + {\amp - 1 \over 2} \tanh( v\, t) \,.
\end{equation}
This resembles a quench occurring around $t=0$ and lasting for a duration $v^{-1}$.
The compact ${\bf S}^1$ provides an energy scale $\ell_c^{-1}$ to the effective 3+1 dimensional theory. At fixed scale factor
(i.e., on Minkowski spacetime), this theory is in a confined phase for temperatures $T < T_c \sim \ell_c^{-1}$ and in a deconfined phase for higher temperatures. We will focus, for simplicity, on the case where the field theory remains in its deconfined phase during the entire evolution. Thus we have some large number  $c_\text{eff}$ of deconfined degrees of freedom potentially participating in the process of particle production in the expanding background.

We will explore how the final energy, entropy, and temperature depend on the initial temperature $T_0$,  the amplitude $\amp$, and the rate $v$ of the expansion.

In the deconfined phase at large $c_\text{eff}$, local 4+1 dimensional quantities do not depend on the scale $\ell_c$, a property known as ``large N volume independence" \cite{Unsal:2010qh}. Holographically this follows from the fact that the finite $\ell_c$ solutions are obtained from the $\ell_c = \infty$ solutions by a trivial identification $x_c \sim x_c + \ell_c$. Using this and scale invariance we we can parameterize the {\em final}  3+1 dimensional energy density in terms of the  dimensionless quantities $\amp$ and $\vT = v/T$ as
\begin{equation}
\label{deff}
\epsilon(\amp,v,T_0) = C \,\ell_c\, v^5 \,f_\epsilon(\amp,\vT) \; .
\end{equation}
where the normalization constant $C$ is related to the equilibrium energy density by $\epsilon_0(T_0) \equiv C\, \ell_c\, T_0^5$. The function $f_\epsilon(\amp,\vT)$ encodes the non-trivial dynamical information.

In the limit $v \to 0$ of adiabatic expansion, the final energy density can be determined from the initial energy density using entropy conservation. The final state is related to the initial state by simple dilution and red shift effects. To isolate the effects of particle creation/energy production, we can compare the final state data relative to that attained simply during the adiabatic expansion for the same starting values. We thus define
\begin{equation}
\label{defF}
F_\epsilon(\amp,\vT) = f_\epsilon(\amp,\vT) - f_\epsilon(\amp, \vT \to 0) \,.
\end{equation}

Alternatively, we can consider the change in temperature, relative to the adiabatic result. Using conformal invariance, we find
\begin{equation}
\label{defG}
T_f = T_f(\amp,\vT \to 0) + v \, \GT(\amp,\vT) \,,
\end{equation}
where the first term gives the adiabatic result, and $\GT(\amp,\vT)$ contains the non-trivial physical information.

Our goal in the remaining sections will be to compute the functions $F_\epsilon(\amp,\vT)$ and
$\GT(\amp,\vT)$ for a holographic theory and for a weakly coupled field theory and compare the results.\footnote{ Since the energy density is simply related to the temperature in a holographic theory, these functions carry the same physical information. However, as we shall describe,  $F_\epsilon$ is more natural at weak coupling, while $\GT$ is a better diagnostic at strong coupling. }

\section{Free Field Theory Results}
\label{sec:}

We begin by analyzing the case of a free conformal field theory, specifically a massless scalar field, in 4+1 dimensions on the background \eqref{FRW}, following \cite{Birrell:1982ix}. We will take $x_c$ non-compact to compare with the strongly-coupled results, which are $\ell_c$-independent.
The scalar action is in general  $d+2$ spacetime dimensions is (nb: we will eventually take $d=3$)\footnote{ We could also consider a massive scalar field in $d+1$ dimensions, instead of compactifying a massless field from $d+2$ dimensions down a Scherk-Schwarz circle. However, we choose to perform the computation in closer analogy with the holographic set-up for ease of comparison. The massive field answer can be recovered by focusing on a single mode in the compact direction, i.e., picking an appropriate value for $k_c$ below.}
\begin{align}
S &= \int d^{d+2} x\; \sqrt{-g} \; \frac{1}{2}\, g^{\mu \nu} \partial_\mu \phi \,\partial_\nu \phi  \,.
\end{align}
Expanding $\phi$ in Fourier modes with ${\bf dk}_d \equiv \frac{d k_c}{2 \pi} \, \frac{d^d k}{(2 \pi)^d}$:
\begin{align}
\phi(x,t) = \int {\bf dk}_d \; \phi_k(t) \; e^{i \left(\vec{k} \cdot \vec{x} + k_c \,x_c\right)} \,,
\end{align}
we obtain a complex harmonic oscillator for each pair $\{(\vec{k},k_c),-(\vec{k},k_c)\}$, with action:
\begin{align}
S = \int dt  \;  a(t)^d\, \bigg\{|\partial_t \phi_k|^2 - \bigg[ \frac{k^2}{a(t)^2} + k_c^2 \bigg]
 |\phi_k|^2 \bigg\} \; .
\label{}
\end{align}
These oscillators are completely decoupled, so their quantum states evolve independently.

\subsection{Time-dependent oscillator}
\label{sec:tdosc}

Classically, the mode with momenta $(\vec{k},k_c)$ evolves as
\begin{equation}
\label{ode}
\ddot{\phi} + d\, {\dot{a} \over a} \dot{\phi} + \left(k_c^2 + {k^2 \over a^2}\right) \phi = 0 \; ,
\end{equation}
For early and late times, this corresponds to a simple harmonic oscillator, with frequencies
$$
\omega_i = \sqrt{k^2 + k_c^2} \qquad \qquad \omega_f = \sqrt{{k^2 \over \amp^2} + k_c^2}
$$
respectively. Defining annihilation operators ${\bf a}$, ${\bf b}$ and ${\bf A}$, ${\bf B}$ associated with the oscillators at early and late times, we can write the field operator at arbitrary times in two equivalent ways
\begin{align}
\phi(t) =
\begin{cases}
{\bf a}\, \phi_i(t) + {\bf b}^\dagger \, \phi^*_i(t) \,,  \\
{\bf A} \,\phi_f(t) + {\bf B}^\dagger \,\phi^*_f(t) \,,
\end{cases}
\label{}
\end{align}
where for $\eta \in \{i, f\}$ one has
\begin{align}
\phi_\eta(t) \to \frac{1}{\sqrt{2\,\omega_\eta}} \, e^{-i\,\omega_\eta\,t} \,, \qquad t \to \pm \infty\,.
\label{eq:inout}
\end{align}
These are normalized such that $ \dot{\phi}_i\,  \phi^*_i - \dot{\phi}^*_i \,\phi_i = i$.

As $\phi_f$ and $\phi^*_f$ form a basis  for solutions to \eqref{ode}, we have (nb:
$|\alpha|^2 - |\beta|^2 = 1$)
\begin{align}
\phi_i(t) = \alpha\, \phi_f(t) + \beta\,  {\phi}^*_f(t) \; .
\label{findBog}
\end{align}
Correspondingly, we have the Bogoliubov transform
\begin{equation}
\begin{split}
{\bf a} &= \alpha^* {\bf A} - \beta^* \,{\bf B}^\dagger  \,, \qquad
{\bf A} = \alpha \,{\bf a} + \beta^*\,  {\bf b}^\dagger \\
{\bf b} &= \alpha^* {\bf B} - \beta^*\, {\bf A}^\dagger \,, \qquad {
\bf B} = \alpha\, {\bf b} + \beta^* \,{\bf a}^\dagger
\end{split}
\label{}
\end{equation}

If we start with an in-mode in the thermal state
\begin{equation}
\rho_T = {1 \over Z} e^{-\beta_T H} = {1 \over Z} e^{-\beta_T\, \omega_i
\left(a^\dagger a + b^\dagger b + 1\right)} \,,
\label{}
\end{equation}
we can determine the final energy in this mode relative to the vacuum energy. One finds
\begin{equation}
\begin{split}
E - E_\text{vac} &= {\rm Tr}\left(\omega_f\, \left({\bf A}^\dagger \,{\bf A} + {\bf B}^\dagger
\,{\bf B}\right)\rho_T\right) \\
&=2 \,\omega_{f} \,\left[ {e^{-\beta_T \omega_{i}} \over  1 - e^{-\beta_T \omega_{i}}} + |\beta|^2 {1 + e^{-\beta_T \omega_{i}} \over  1 - e^{-\beta_T \omega_{i}}} \right]
\end{split}
\label{eq:enprodphi}
\end{equation}
Here, the first term corresponds to the adiabatic result. We can interpret it as having the same occupation probabilities as the initial state, but with modified energies. The second term, which vanishes in the limit of slow expansion where the Bogoliubov coefficient $\beta$ goes to zero, captures the effects of particle creation.

\subsection{Results}
\label{sec:Free_results}

\begin{figure*}[t]
\begin{subfigure}{0.49\textwidth}
\includegraphics[width=3in]{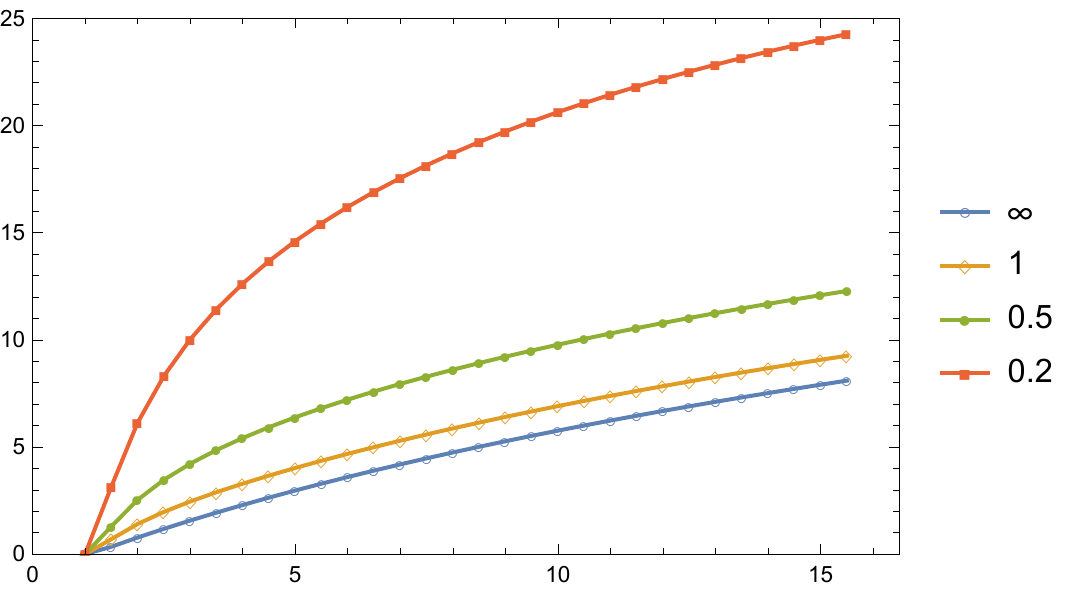}
 \begin{picture}(0,0)(0,0)
 \put(180,15){\makebox(0,0){$\amp$}}
 \put(-5,120){\makebox(0,0){$F_\epsilon$}}
 \put(100,75){\makebox(0,0){$\downarrow \; \vT$}}
 \put(15,135){\makebox(0,0){$^{^{\times 10^{-3}}}$}}
 \end{picture}
\subcaption{}
\label{fig:FvAfree}
\end{subfigure}
\hfill
\begin{subfigure}{0.49\textwidth}
\includegraphics[width=3in]{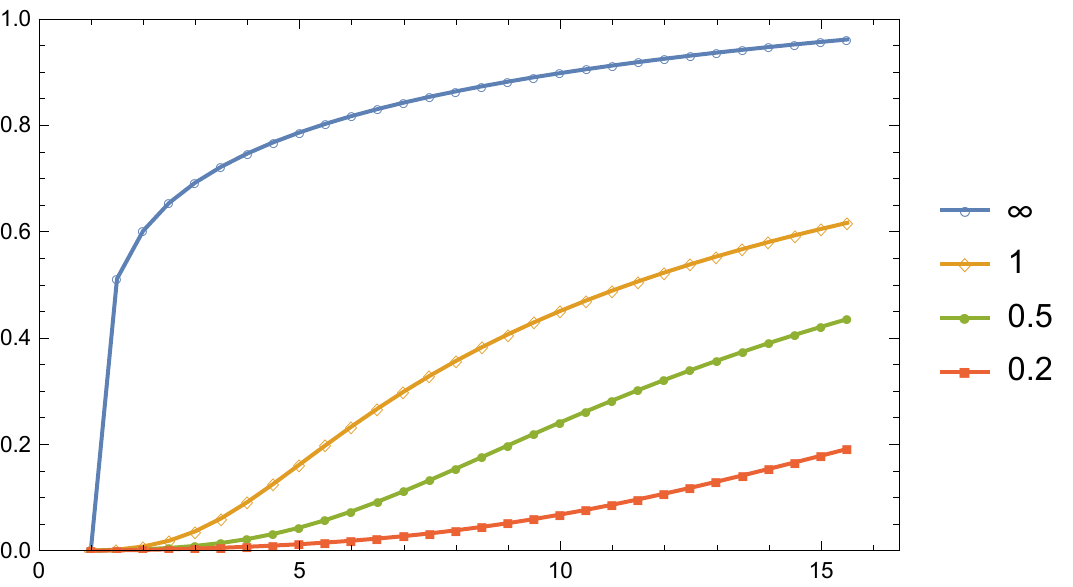}
\begin{picture}(0,0)(0,0)
\put(180,15){\makebox(0,0){$\amp$}}
\put(-5,120){\makebox(0,0){$\GT$}}
\put(100,75){\makebox(0,0){$\uparrow \; \vT$}}
\end{picture}
\subcaption{}
\label{fig:GvAfree}
\end{subfigure}
\caption{Free field results for: (a) the energy production captured by  $F_\epsilon(\amp,\vT)$ and
(b) the temperature change  $\GT(\amp,\vT)$, as function of the amplitude for a range of
 $\vT$ as indicated.
This should be contrasted with the strong coupling result displayed in Fig.~\ref{fig:holFG}. }
\label{fig:freeFG}
\end{figure*}
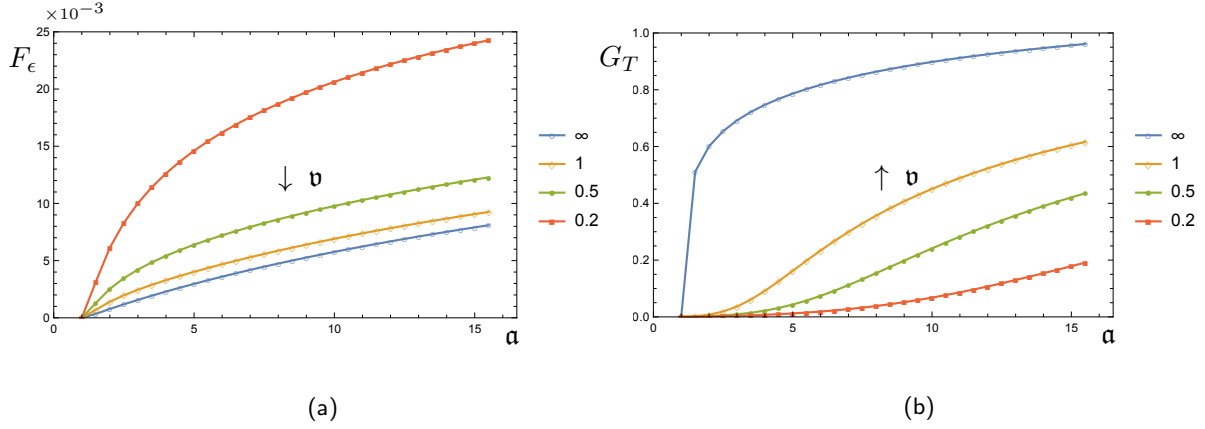

We can now add the energies for all the modes of the scalar field to obtain an expression for the final energy density.  The density of modes in $k$ space with an IR regulator $L$ is $\left(\tfrac{L}{2 \pi}\right)^{d+1}$, while the  total volume after expansion is $L^{d+1} \,\amp^d$. The total energy density is then the integral over all $(k,k_c)$ of \eqref{eq:enprodphi}. Normalizing the result correctly by the mode density and the final proper volume, we find the final answer (independent of the IR regulator) to be:\footnote{Since we integrate over ${\bf k}$ and $-{\bf k}$ independently, we have to divide the result of \eqref{eq:enprodphi} by half. }
\begin{align}
\epsilon_E(\amp,v,T) = \frac{1}{\amp^d} \int {\bf dk}_d\; \frac{E-E_\text{vac}}{2} \,.
\label{finalE}
\end{align}

For a $d=3$ free scalar  the initial energy density (equivalently, the result for $\amp=0$) is
\begin{align}
\label{TEfree}
\epsilon_0(T_0) = C_\phi\,\ell_c\, T_0^5\,, \quad C_\phi =\frac{3\, \zeta(5)}{\pi^2}.
\end{align}
Thence  from \eqref{defF} and \eqref{deff}, we find
\begin{align}
F_\epsilon(\amp,\vT) = \frac{1}{C_\phi\,\amp^3} \int {\bf dk}_3\; \omega_f\;  |\beta|^2_{v=1}\,
\frac{ e^{\vT\, \omega_i} +1}{ e^{ \vT\,\omega_i}-1} \; .
\label{eq:Ffree}
\end{align}
To evaluate $F_\epsilon$, we calculate the Bogoliubov coefficient $\beta(v,k,k_c,\amp)$ numerically by finding solutions of \eqref{ode}.  We pick initial conditions $\phi_i =e^{-i \omega_i t}$ for early times and read off $\beta$ from the decomposition \eqref{findBog} in the late time behavior. Our results for the function $F_\epsilon$ for various choices of $\vT$ are shown in Fig.~\ref{fig:FvAfree}.

Similarly, we can calculate the function $\GT(\amp,\vT)$ defined in \eqref{defG}. Note that in the free field case, the final state is not thermal. However, assuming some very weak interactions that thermalize the system on a time scale much longer than the expansion time, we can define the final temperature in terms of the final energy density using the equilibrium relation \eqref{TEfree}. Our results for the function
$\GT(\amp,\vT)$ are shown in Fig.~\ref{fig:GvAfree}.

\section{Holographic Strong Coupling Results}
\label{sec:setup}

We now derive the corresponding results for the case of a strongly-coupled holographic field theory. For this purpose, we seek asymptotically \AdS{6} solutions to Einstein's equations with negative cosmological constant, with boundary geometry given to be the FRW cosmology \eqref{FRW}.\footnote{ For the strong coupling calculations, we can easily generalize to allow spatial curvature, replacing $d\vec{x}^2 \to ds_{3,\kappa}^2$, for spatial slices with spatial curvature $\kappa=0,\pm1$. However, we will mostly discuss the $\kappa=0$.}

The bulk metric, written in ingoing Eddington-Finkelstein or Bondi-Sachs coordinates, is ($\ell_\text{AdS}=1$):
\begin{align}
&ds^2 =-2 A\, e^{2 \chi}\,dt^2+2  e^{2 \chi}\,dt \,dr +\Sigma^2 (e^B \, d{\vec x}^2+e^{-3B} \,dx_c^2)
\label{model2}
\end{align}
Since we are interested in homogeneous cosmologies, we take the metric functions $\{A,\chi,\Sigma,B\}$ to depend in the holographic radial direction $r$ and the time coordinate $t$, but to be independent
of the spatial coordinates $\{ \vec{x}, x_c\}$.
The boundary scale factor behaves as in \eqref{timedep}.

In order to solve the Einstein equations we start the system at an initial thermal state.  The final state of the system will be also be thermal (even starting from vacuum), and we can read off the final energy density, entropy, and temperature of the final state from the solution at late times. We can then calculate the quantities $F_\epsilon(\amp,\vT)$ and $\GT(\amp,\vT)$ for comparison with the free field results. Details of the numerical calculation are given in the Appendix.

\subsection{Equilibrium and Adiabatic Physics}
\label{sec:initial}

To compute $F_\epsilon$ and $\GT$, we need the equilibrium results for the holographic theory.  In the deconfined phase,  the CFT$_{4+1}$ compactified on a spatial circle is dual to the Schwarzschild-\AdS{6} solution compactified on ${\bf S}^1_c$, from which we can read off the thermodynamic data:
\begin{equation}
\begin{split}
 \epsilon   &= 4\, p = 4\, C_H \, T^5\,,  \qquad s = \frac{dp}{dT} \,,  \\
  C_H &=c_\text{eff} \left(\frac{4\pi}{5}\right)^5.
\end{split}
\label{equil}
\end{equation}
The physical quantities on the FRW$_4$ universe are obtained simply by multiplying these expressions by
$\ell_c$. For $\kappa \neq 0$ we can obtain the equilibrium thermodynamics numerically. Furthermore, for an adiabatic expansion from an initial scale factor $a_i=1$ to a final scale factor $a_f = \amp$, keeping  $\ell_c$ fixed, the total entropy should remain fixed. Hence from \eqref{equil} we obtain

\begin{equation}
\label{adiabatic}
T_f = {1 \over \amp^{3 \over 4}} \,  T_i  \; \Longrightarrow\;  {\langle T_{00} \rangle_f \over \langle T_{00} \rangle_i }= \left( {T_f \over T_i} \right)^5 = {1 \over \amp^{15 \over 4}}
\end{equation}
as the proper volume of the FRW$_4$ scales as $a(t)^3$. This dilution factor in energy, which is determined by the underlying scale invariance of CFT$_{4+1}$,  is part way between that for matter and that for radiation. This is sensible, since we should have a factor of $\amp^{-3}$ for the volume dilution, but the red shift effect operates only in 3 of the 4 spatial directions.

\subsection{Results}
\label{sec:strong}

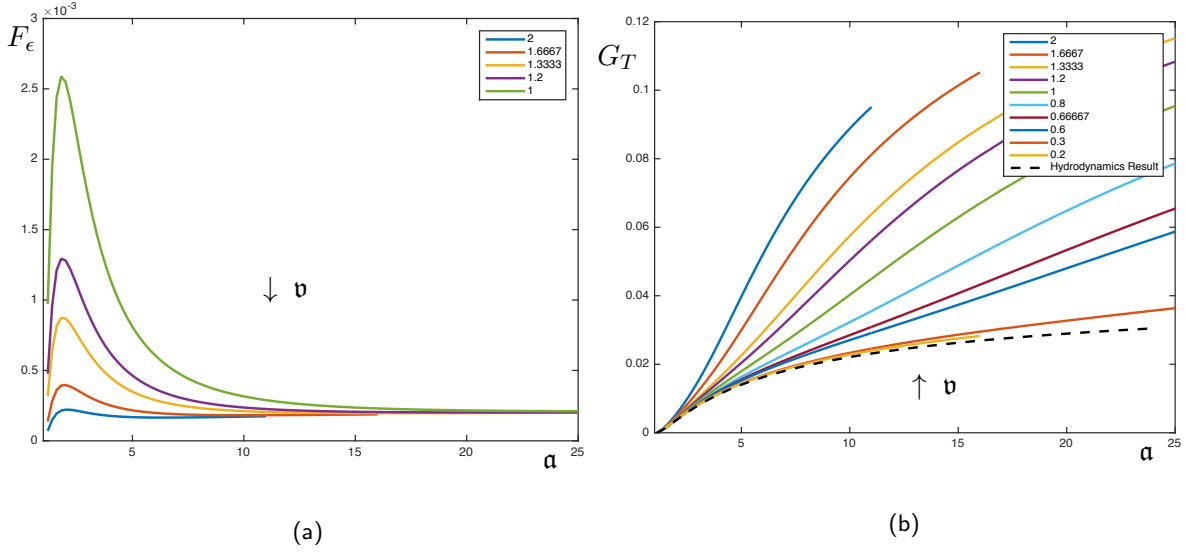
\begin{figure*}[t]
\begin{subfigure}{0.49\textwidth}
\includegraphics[width=3in]{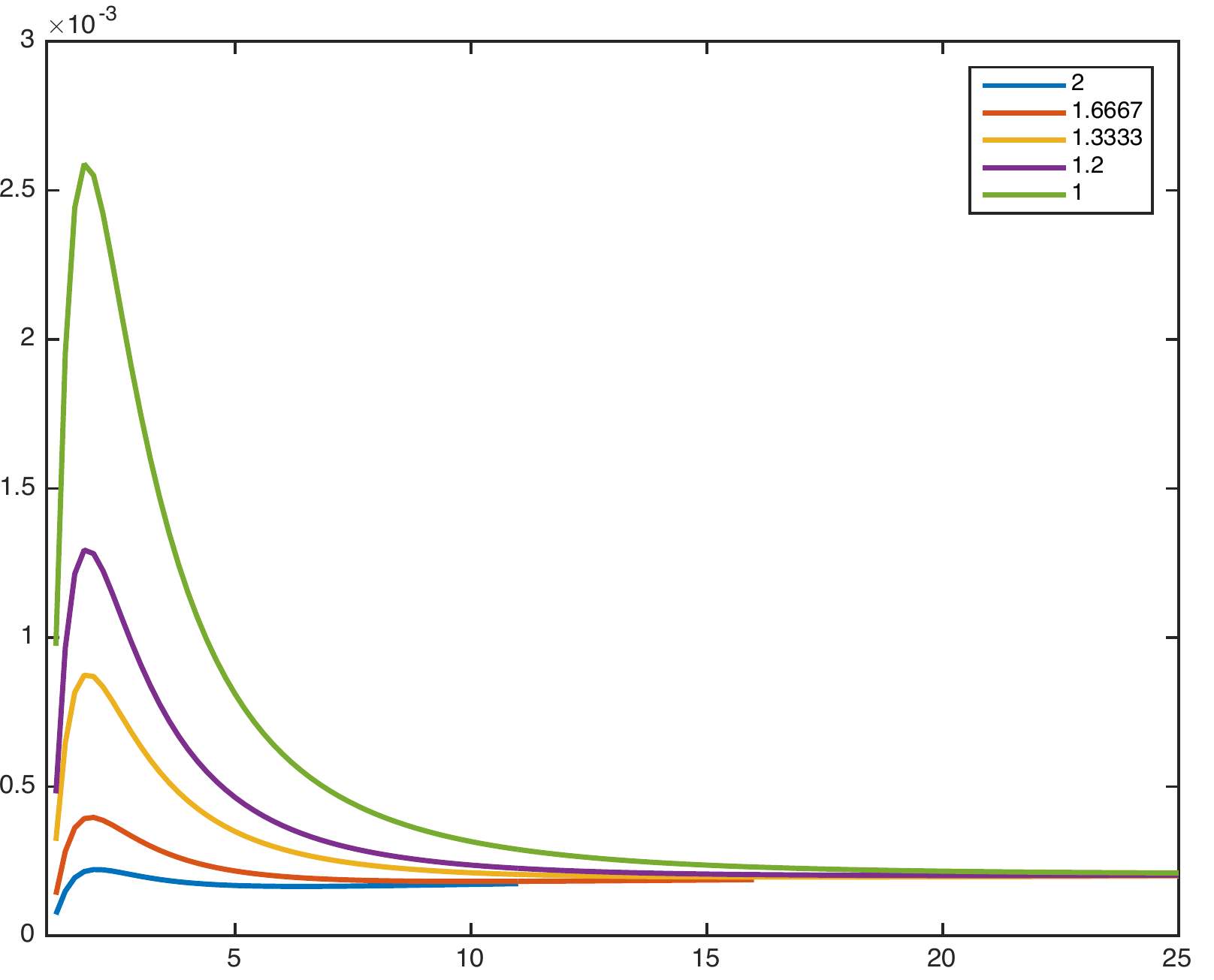}
 \begin{picture}(0,0)(0,0)
 \put(200,15){\makebox(0,0){$\amp$}}
 \put(0,175){\makebox(0,0){$F_\epsilon$}}
 \put(100,80){\makebox(0,0){$\downarrow \; \vT$}}
 \end{picture}
\subcaption{}
\label{fig:FvAs}
\end{subfigure}
\hfill
\begin{subfigure}{0.49\textwidth}
\includegraphics[width=3in]{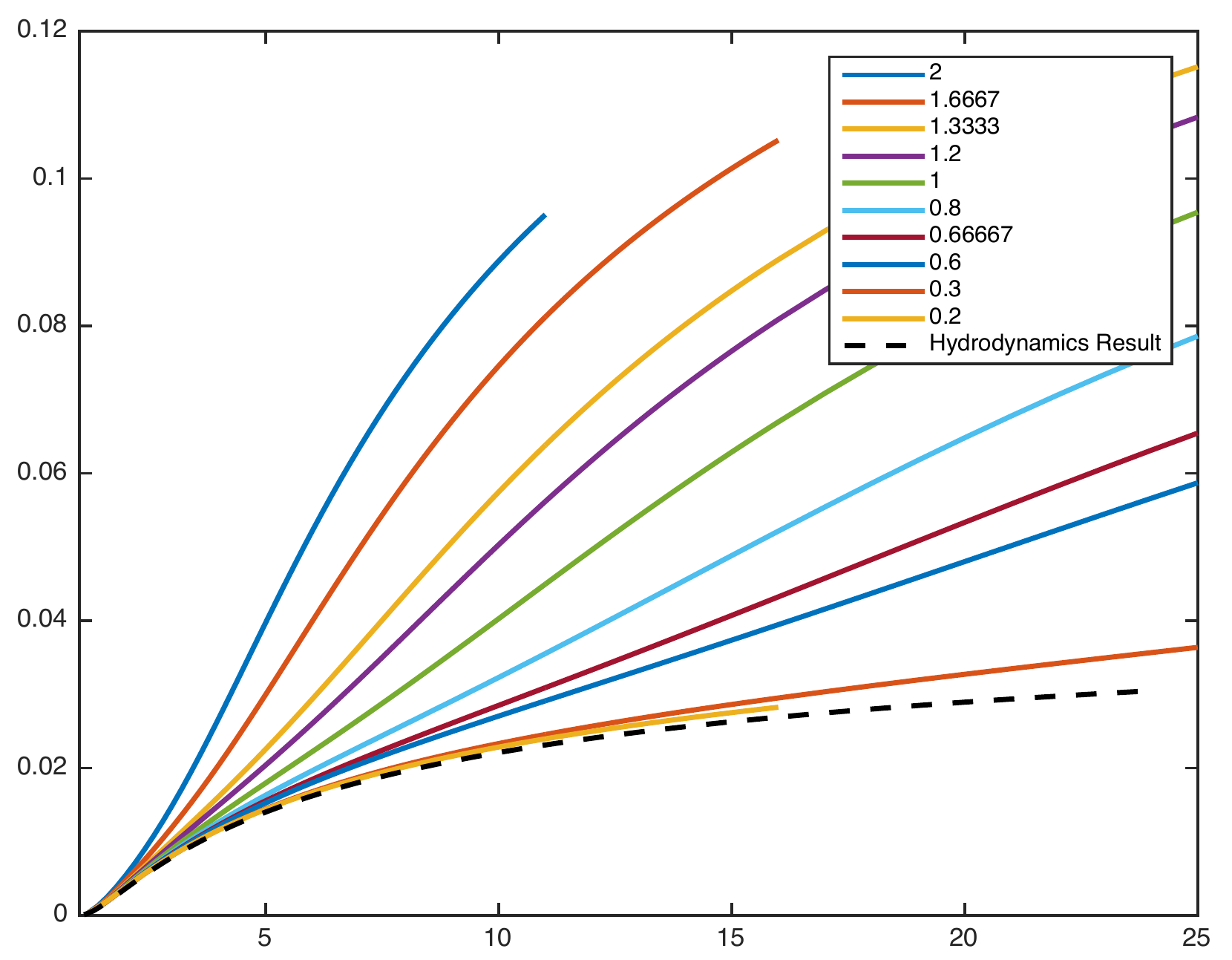}
 \begin{picture}(0,0)(0,0)
 \put(200,15){\makebox(0,0){$\amp$}}
 \put(0,165){\makebox(0,0){$\GT$}}
 \put(120,40){\makebox(0,0){$\uparrow \; \vT$}}
 \end{picture}
\subcaption{}
\label{fig:GvAs}
\end{subfigure}
\caption{Holographic results at strong coupling as function of the amplitude for a range of
 $\vT$. Contrast against Fig.~\ref{fig:freeFG} for free field analog.
 \\
(a) The energy production captured by  $F_\epsilon(\amp,\vT)$. This quickly becomes insensitive to the initial state as amplitude is increased. \\
(b) The temperature change  $\GT(\amp,\vT)$, displaying good agreement with the hydrodynamics prediction (dashed line) for small enough $\vT$.
 }
\label{fig:holFG}
\end{figure*}

Using the equilibrium \eqref{equil} and adiabatic \eqref{adiabatic} results, we have calculated
$F_\epsilon(\amp,\vT)$ and $\GT(\amp,\vT)$ numerically for various values of $\vT$; see Fig.~\ref{fig:holFG}.
From Fig.~\ref{fig:FvAs}, we note that as the amplitude of expansion increases, the function $F_\epsilon(\amp,\vT)$ describing the final energy density approaches the same approximately constant curve for each value of the initial temperature (or $\vT$), in contrast to the free field theory results. This suggests that at strong coupling, the field theory quickly forgets about the initial state before the expansion, with the energy density in the final state sensitive only to the expansion rate.

For $G(\amp,\vT)$, we also find a convergence of the results for various values of $\vT$, but this time for sufficiently small $\amp$., cf., Fig.~\ref{fig:GvAs}. As we will see now, this may be understood as a consequence of the fact that in this regime, the strongly-coupled field theory dynamics is well-described by relativistic hydrodynamics. Generally, we expect this to be true in the case that derivatives are small compared to the scale set by the temperature. In our homogeneous setup, physical quantities vary only in time, and the scale of this time dependence is characterized by the Hubble parameter $H = \dot{a}/a$. Thus, we expect hydrodynamics to be valid when $H \ll T$. In this fluid-dynamical regime, the field theory stress tensor is determined by the local temperature $T(x)$ and velocity $u^\mu(x)$, see \cite{Hubeny:2011hd}:
\begin{equation}
\begin{split}
T_{\mu \nu} &=C_H\, T^4 \bigg[ T\, (g_{\mu\nu} + 5\, u_\mu\, u_\nu)  \\
& - \frac{5 }{2 \pi} \,\left( P_{\mu\alpha} P_{\nu\beta} -\frac{1}{4}\, P_{\mu \nu} P_{\alpha\beta}\right) \nabla^\alpha u^\beta \bigg]+ {\cal O}(\nabla^2) \; ,
\end{split}
\label{}
\end{equation}
where $P_{\mu \nu} = g_{\mu \nu} + u_{\mu} u_{\nu}$ defines a projection to the spatial directions. The dynamics of $T(x)$ and $u^\mu(x)$ are then determined by the conservation relation $\nabla^\mu T_{\mu \nu} = 0$. In our case, the spatial velocity vanishes, so $u^\mu = \delta^\mu_t$. The evolution equation  is simply
\begin{align}
\dot{T} + \frac{3}{4} \, T \, \frac{\dot{a}}{a} = \frac{3}{32 \pi} \,\left(\frac{\dot{a}}{a} \right)^2
\label{eq:Thydro}
\end{align}
Starting from initial temperature $T_0$ with scale factor (\ref{timedep}), we find the solution
$\GT(\amp,\vT) = T(t=\infty) - \frac{T_0}{\amp^\frac{3}{4}}$:
\begin{align}
\GT(\amp,\vT) = \frac{1}{7 \pi}  \; \frac{\amp^\frac{7}{4} + 7 \,\amp^\frac{3}{4} - 7 \,\amp - 1}{\amp^\frac{3}{4} (\amp-1)}
\label{}
\end{align}

The salient feature is that $\GT$ is independent of $\vT$ in the hydrodynamic regime. From the numerical results, Fig.~\ref{fig:GvAs}, we  see  very good agreement  with this prediction for small $\vT$, just as expected from the criterion $H \ll T$. This serves as a strong check of the numerical methods.
Note that for large enough temperatures the holographic results are well-described by hydrodynamics for an increasing range of amplitudes. At lower temperatures, and for sufficiently large amplitudes, the non-linear evolution deviates strongly from hydrodynamics. It is curious that such deviations are always positive, i.e., lead to more energy and entropy production than in hydrodynamics.

\section{Discussion}
\label{sec:discussion}

In this paper, we have presented results for the evolution of homogeneous states of strongly-coupled confining gauge theories in FRW cosmologies. While we have focused on the flat $k=0$ case, explicit results have also been obtained in the case of positive ($k=1$) and negative ($k=-1$) spatial curvature. We leave this exploration to future study.

We have found significant qualitative differences from the case of free field theory. First, for the range of temperatures considered, the final energy density quickly becomes insensitive to the initial temperature as the amplitude of the expansion becomes large (for fixed expansion rate). Second, the small amplitude results are well described by a hydrodynamic approximation in which the difference $T_f - T_f\big|_\text{adiabatic}$ is independent of the initial temperature.

Using similar techniques, it would be straightforward to consider other cosmological spacetimes, or to compute other observables such as correlation functions and entanglement entropies.
A more challenging line of research would be to study the time-dependent deconfinement phase transition, caused by the expansion.

\hspace{5mm}
\begin{acknowledgments}

M.~Rangamani and M.~Van Raamsdonk would like to thank KITP, Santa Barbara for hospitality during the concluding stages of the project, where their work was supported in part by the National Science Foundation under Grant No. NSF PHY11-25915.
M.~Rangamani was supported in part by the STFC Consolidated Grant ST/L000407/1, and by the ERC Consolidator Grant Agreement ERC-2013-CoG-615443: SPiN. M.~Rozali is supported by a Discovery Grant from NSERC of Canada. M. ~Van Raamsdonk is supported by NSERC and FQXi.

\end{acknowledgments}

\appendix

 \section{Appendix: Solution Method}
 \label{sec:supplement}

\noindent
{\bf Equations Of Motion:} In choosing the Bondi-Sachs form of the metric (\ref{model2}), we are able to use the characteristic formulation of Einstein equations. The numerical scheme we use is described in great detail in \cite{Chesler:2013lia}, and some of our choices are more similar to the ones made in \cite{Balasubramanian:2013yqa}. Here we briefly comment on some issues specific to the models discussed here. We refer the interested reader to \cite{Chesler:2013lia,Balasubramanian:2013yqa}
for a more complete discussion.

 The Einstein equations, in the nested form used for their solution, are as follows:
\begin{align}
\frac{1}{4}\, \Sigma \, B'^2- 2\, \Sigma' \, \chi '+\Sigma''=0
\nonumber
\end{align}
where prime indicates a radial  ($r$) derivative. We choose to gauge fix the determinant of the spatial metric $\Sigma$ and solve for the field $\chi$ at each time step. The gauge fixing is not complete: we leave enough freedom to fix the coordinate location of the apparent horizon, for convenience. The gauge freedom is implemented in terms of a gauge parameter $\lambda(t)$ which we treat as a dynamical variable.

Next, we solve for the field $d_{+}\Sigma$, where $d_{+}=\partial_t+A \,\partial_r$:
\begin{align}
d_{+}\Sigma'+\frac{d_{+}\Sigma\; \Sigma'}{\Sigma}-\frac{3}{2}\, e^{2 \chi}\, \Sigma=0
\label{}
\end{align}
Since the field $\Sigma$ is gauge fixed, leaving only the parameter it $\lambda$ in its stead,  $d_{+}\Sigma$ should be thought of as a proxy for the field $A$ and the time derivative of the gauge parameter $\dot{\lambda}$, which we solve for later after gathering more information (dot denotes time derivative).

The next equation is for the  derivative of the dynamical field B:
\begin{align}
d_{+}B'+\frac{d_{+}B \; \Sigma'}{\Sigma}+\frac{d_{+}\Sigma \;B'}{\Sigma}=0
\end{align}
Once this is solved, we can find   $\dot{\lambda}$ by the requirement that the apparent horizon (defined as the locus of the outermost zero of $d_{+} \Sigma$), stays at fixed radial location
\begin{align}
A+\frac{1}{6}\;e^{-2 \chi}\; d_{+}B^2=0
\nonumber
\end{align}

When $\dot{\lambda}$ is obtained, the expression for $d_{+}\Sigma$ is sufficient to find the field $A$, and that information in turn can be used to convert knowledge of $d_{+}B$ into an expression for the time derivative   $\dot{B}$.

The above nested scheme can be used for constrained evolution: at each time step we are given the values of the propagating fields $\{B,\lambda\}$. We use the above process to determine the constrained fields $\{\chi, A\}$ as well as the time derivatives of the dynamical fields, $\{\dot{B},\dot{\lambda}\}$, which are then used to propagate them to the next time step.

\medskip
\noindent
{\bf Asymptotic Expansion and Observables:} Denote the metric elements in our ansatz collectively as $g_{ij} (t,r)$. For each such metric element, we can write an asymptotic near-boundary expansion, by transforming the familiar Fefferman -Graham expansion to incoming coordinates. Since in our case (AdS$_{d+1}$ with d odd) there are no logarithmic terms, the expansion is simple:\footnote{ A similar expansion in situations with $d$ even would lead to logarithmic terms which would appear at order $\frac{1}{r^d}\,\log(r)$.}
\begin{align}
g_{ij}(t,r)=r^2\,\left(g^{(0)}_{ij}(t)+\frac{g^{(1)}_{ij}(t)}{r}+\ldots+\frac{g^{(d)}_{ij}(t)}{r^d}\right)+ \cdots\,.
\nonumber
\end{align}

\medskip
\noindent
\begin{figure}
\centering
\vspace{-2cm}
\includegraphics[width=8cm,height=10cm]{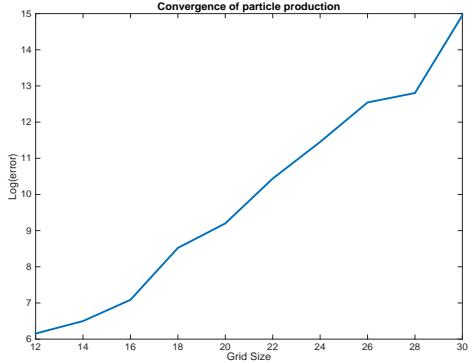}
\vspace{-2.5cm}
\caption{Convergence of the total entropy production with the grid size, used to discretize the radial direction. Plotted is the (base 10) logarithmic change of total entropy as we refine the grid, for fixed parameters. This demonstrates the expected exponential convergence. The plots in the paper were generated with 30 grid points, where the error corresponds roughly to machine precision. }
\label{fig:converg}
\end{figure}

As is well known, all the expansion coefficients up to, but not including $g^{(d)}_{ij}(t)$, are determined in terms of the boundary metric $g^{(0)}_{ij}(t)$. Since  the coefficients are known, we choose to shift the fields by these expressions, and solve for the shifted dynamical fields. Such choice of variables is described in \cite{Chesler:2013lia}, but in our case the shift includes all the expansion coefficients determined by the boundary metric. As a practical matter, since the expressions for the expansion coefficients are increasingly complex as the dimensionality increases, the resulting equations and boundary conditions are extremely long and uninformative, and we spare the reader the precise details.

To extract the energy-momentum tensor from the asymptotic expansion we use the standard expressions for the counter-terms and renormalized energy momentum tensor, given e.g., in \cite{deHaro:2000xn}.

\medskip
\noindent
{\bf Equilibrium thermodynamics:} To illustrate the above, consider the derivation of the equilibrium thermodynamic data quoted in \eqref{equil}. The Schwarzschild-\AdS{6} solution with a flat boundary metric ($\kappa=0$) is given by:
\begin{align}
ds^2 &= {1 \over z^2} \left(-f(z) \,dt^2 +dz^2+ g(z) \,(ds_{3,0}^2 + d w^2)\right)
\nonumber \\
f(z) &= {(1- \frac{1}{4}\, \mu \, z^5)^2 \over (1 +\frac{1}{4} \,\mu\, z^5)^{6 \over 5}} \,,\qquad
g(z) = (1 +\frac{1}{4}\, \mu \, z^5)^{4 \over 5}
\end{align}
The CFT$_5$ stress tensor is given by $\langle T_{\mu \nu} \rangle = {5 \over 16 \pi G_N} \Gamma^{(5)}_{\mu \nu}$,
where $\Gamma^{(5)}_{\mu \nu}$ is the $z^5$ term in the asymptotic expansion of the metric. Setting $c_\text{eff} = \frac{1}{16\pi G_N}$ as the central charge of the CFT and integrating over the compact ${\bf S}^1_w$ we have
\begin{equation}
\langle T_{00} \rangle =4\, c_\text{eff}\,  \mu \,, \qquad
\langle T_{ij} \rangle = c_\text{eff}\, \mu \, \delta_{ij} \,.
\label{}
\end{equation}
The entropy density $s$  computed from the horizon area using the Bekenstein-Hawking formula  $S = \frac{1}{4\,G_N}$, gives $s = 4\pi\, c_\text{eff} \,\mu^\frac{4}{5} $ (nb: horizon is located at the zero locus $f(z_+) =0$). The temperature of the black hole (say using $dE = T\, dS$) is obtained to be $ T = \frac{5}{4\pi} \, \mu^{1 \over 5}$.

Note that one of the advantages of working in \AdS{6} is that we do not have to worry about logarithmic terms in the asymptotic expansion, which plague odd-dimensional AdS spacetimes.

\noindent
{\bf Numerical Choices:} The solution method consists of constrained evolution, which involves the iterative solution of several linear ordinary differential equations at each time step. We discretize those differential equations using pseudo-spectral collocation methods. To evolve the system in time we use a fourth order Runge-Kutta method with an adaptive step size.
In order
to avoid numerical instabilities we use filtering at each time
step; we use both filters based on Chebyshev interpolation, or
filters based on fast Fourier transform.

Since the expressions involved in the equations are quite long, we need to employ some special tricks to minimize round-off errors. To this end we use compensated summation to evaluate the sums involved in our equations, and iterative refinement in solving the linear equations. Both those steps utilize extra precision in intermediate steps of the calculation.

We demonstrate the convergence of our solution in
Fig.~\ref{fig:converg}. Similar tests were performed for other numerical parameters, such as the tolerance involved in determining the temporal step size.



\begin{thebibliography}{1}

\bibitem{Marolf:2013ioa}
D.~Marolf, M.~Rangamani, and T.~Wiseman, {\it {Holographic thermal field theory
  on curved spacetimes}},  {\em Class.Quant.Grav.} {\bf 31} (2014) 063001,
  [\href{http://xxx.lanl.gov/abs/1312.0612}{{\tt arXiv:1312.0612}}].

\bibitem{Chesler:2008hg}
P.~M. Chesler and L.~G. Yaffe, {\it {Horizon formation and far-from-equilibrium
  isotropization in supersymmetric Yang-Mills plasma}},  {\em Phys.Rev.Lett.}
  {\bf 102} (2009) 211601, [\href{http://xxx.lanl.gov/abs/0812.2053}{{\tt
  arXiv:0812.2053}}].

\bibitem{Hawking:2000bb}
S.~W. Hawking, T.~Hertog, and H.~S. Reall, {\it {Trace anomaly driven
  inflation}},  {\em Phys. Rev.} {\bf D63} (2001) 083504,
  [\href{http://xxx.lanl.gov/abs/hep-th/0010232}{{\tt hep-th/0010232}}].

\bibitem{Unsal:2010qh}
M.~Unsal and L.~G. Yaffe, {\it {Large-N volume independence in conformal and
  confining gauge theories}},  {\em JHEP} {\bf 1008} (2010) 030,
  [\href{http://xxx.lanl.gov/abs/1006.2101}{{\tt arXiv:1006.2101}}].

\bibitem{Birrell:1982ix}
N.~Birrell and P.~Davies, {\it {Quantum Fields in Curved Space}},  {\em
  Cambridge Monogr.Math.Phys.} (1982).

\bibitem{Hubeny:2011hd}
V.~E. Hubeny, S.~Minwalla, and M.~Rangamani, {\it {The fluid/gravity
  correspondence}},  in {\em {Black holes in higher dimensions (ed. G. Horowitz)}}, {\em Cambridge U. Press}, pp.~348--383, (2012) [\href{http://xxx.lanl.gov/abs/1107.5780}{{\tt arXiv:1107.5780}}].


\bibitem{Chesler:2013lia}
P.~M. Chesler and L.~G. Yaffe, {\it {Numerical solution of gravitational
  dynamics in asymptotically anti-de Sitter spacetimes}},  {\em JHEP} {\bf
  1407} (2014) 086, [\href{http://xxx.lanl.gov/abs/1309.1439}{{\tt
  arXiv:1309.1439}}].

\bibitem{Balasubramanian:2013yqa}
K.~Balasubramanian and C.~P. Herzog, {\it {Losing Forward Momentum
  Holographically}},  {\em Class.Quant.Grav.} {\bf 31} (2014) 125010,
  [\href{http://xxx.lanl.gov/abs/1312.4953}{{\tt arXiv:1312.4953}}].

\bibitem{deHaro:2000xn}
S.~de~Haro, S.~N. Solodukhin, and K.~Skenderis, {\it {Holographic
  reconstruction of space-time and renormalization in the AdS / CFT
  correspondence}},  {\em Commun.Math.Phys.} {\bf 217} (2001) 595--622,
  [\href{http://xxx.lanl.gov/abs/hep-th/0002230}{{\tt hep-th/0002230}}].

\end{thebibliography}

\providecommand{\href}[2]{#2}\begingroup\raggedright\endgroup


\end{document}